\documentclass[journal]{IEEEtran}
 \usepackage{times, mathptm, graphicx, amsmath, amsfonts, bm, mathtools}
 \def\DS{\displaystyle}
 \def\TS{\textstyle}
 \def\BEA{\begin{eqnarray}}
 \def\EEA{\end{eqnarray}}
 \def\NN{\nonumber}
 
 \def\IE{{\it i.e.}}
 \def\EG{{\it e.g.}}
 \def\ETC{{\it etc.}}
 \def\DEF{:=}
 
 \def\VEC#1{{\pmb{#1}}}
 \def\REAL{{\bf R}}
 \newtheorem{definition}{\bf Definition}
 \newtheorem{statement}{\bf Statement}
 \renewcommand{\leftmark}{INSTANTONS CAUSING ITERATIVE DECODING TO CYCLE}
 \DeclareMathOperator*{\argmin}{arg\,min}

\begin{document}

\title{Instantons causing iterative decoding to cycle}

\author{Misha Stepanov \thanks{This work was supported by NSF grant 
DMS-0807592 ``Asymptotic Performance of Error Correcting Codes''.} 
\thanks{M.~Stepanov is with Department of Mathematics and Program in 
Applied Mathematics, University of Arizona, Tucson, AZ 85721, USA 
(e-mail: stepanov@math.arizona.edu).}}

\maketitle

 \begin{abstract}
  It is speculated that the most probable channel noise realizations 
(instantons) that cause the iterative decoding of low-density 
parity-check codes to fail make the decoding not to converge. The 
Wiberg's formula is generalized for the case when the part of a 
computational tree that contributes to the output at its center is 
ambiguous. Two methods of finding the instantons for large number of 
iterations are presented and tested on Tanner's $\pmb{[155, 64, 20]}$ 
code and Gaussian channel. The inherently dynamic instanton with 
effective distance of $\pmb{11.475333}$ is found.
 \end{abstract}


\section{Introduction}

Low-density parity-check (LDPC) codes \cite{63Gal, 1996_EL_MN, Wiberg} 
with iterative decoding got a lot of attention due to their excellent 
performance. The decoding error probability is larger than one could 
expect when the Signal-to-Noise Ratio (SNR) is high, a phenomenon called 
error floor \cite{2003_ENTCS_MP, 03Rich}.

In some cases the substructures of the code that provide a leading 
contribution to the error probability are known: they are {\it 
codewords} in the case of maximum likelihood decoding, and {\it stopping 
sets} \cite{2002_IEEETIT_DPTRU} in the case of iterative decoding and 
binary erasure channel. For general situation several heuristics were 
introduced: {\it near-codewords} \cite{2003_ENTCS_MP} or {\it trapping 
sets} \cite{03Rich} as bits subsets that violate just a few parity 
checks, {\it pseudo-codewords} \cite{Wiberg, 2007_IEEETIT_KS} as the 
codewords on {\it computational tree}, {\it pseudo-codewords} as 
non-codeword vertices of a polytope used in linear programming decoding 
\cite{2005_IEEETIT_FWK}, {\it [fully] absorbing sets} 
\cite{2009_IEEETWC_ZDNAW}, and {\it instantons} \cite{2005_PRL_SCCV}. 
Even if the description of the deleterious substructures is available, 
it still could be a non-trivial problem to find them.

The instanton-amoeba scheme \cite{2005_PRL_SCCV,ISIT2006_SC} while being 
quite effective in getting instantons for small number of iterations 
$n_{\rm iter}$ (with about $10$ iterations being the maximum in 
practice) is having difficulties in finding the instantons for large 
$n_{\rm iter}$. This work suggests two new methods to find the most 
probable channel noise configurations that cause the iterative decoder 
[with large number of iterations] to fail.

LDPC codes can be defined by parity check matrix ${\hat H}$ or Tanner 
graph \cite{1981_IEEETIT_T} which is a sparse bipartite graph with two 
sets of vertices: bits and parity checks. The notation $i 
\circ\hskip-2.7pt-\hskip-2.3pt\raisebox{-0.9pt}{\mbox{\scalebox{0.95}{\rotatebox{45}{$\diamond$}}}} 
\alpha$ is used to indicate that $H_{\alpha i} = 1$ and the bit $i$ and 
the check $\alpha$ are connected by an edge.

The binary (made of $+1$ and $-1$ numbers, or just ``$+$''s and 
``$-$''s) codeword $\VEC{\sigma} = (\sigma_1, \sigma_2, \dots, 
\sigma_N)$ is transmitted over a noisy channel with continuous output 
$\VEC{x} = (x_1, x_2, \dots, x_N)$. In the paper the channel is assumed 
to be memoryless, \IE, $P(\VEC{x} | \VEC{\sigma}) = \prod_{i = 1}^N 
P(x_i | \sigma_i)$. The decoder takes the logarithmic likelihoods $h_i = 
(1 / 2) \log \big( P(+| x_i) / P(-| x_i) \big)$ at each bit $i$ as an 
input, where $P(\pm| x) = P(x| \pm) / \big( P(x|+) + P(x|-) \big)$.

The only iterative decoding that is used in the paper is the min-sum 
algorithm
 \begin{align*} {\displaystyle\mbox{decoding output:~~}} & m_i^{(k)} = 
    h_i
+ \sum_{\alpha}^{i 
\circ\hskip-1.25pt-\hskip-1.7pt\raisebox{-0.8pt}{\scalebox{0.75}{\mbox{\rotatebox{45}{$\diamond$}}}} 
\alpha} \mu_{\alpha \to i}^{(k - 1/2)} \NN \\
    {\displaystyle\begin{array}{c} \raisebox{-1pt}{\mbox{bits $\to$ 
checks}} \\ \raisebox{1pt}{\mbox{messages}} \end{array}\!\!\!\!:~~} & 
\eta_{i \to \alpha}^{(k)} = h_i + \sum_{\beta \ne \alpha}^{i 
\circ\hskip-1.25pt-\hskip-1.7pt\raisebox{-0.8pt}{\scalebox{0.75}{\mbox{\rotatebox{45}{$\diamond$}}}} 
\beta} \mu_{\beta \to i}^{(k - 1/2)} \NN \\
    {\displaystyle\begin{array}{c} \raisebox{-1pt}{\mbox{checks $\to$ 
bits}} \\ \raisebox{1pt}{\mbox{messages}} \end{array}\!\!\!\!:~~} & 
\mu_{\alpha \to i}^{(k + 1 / 2)} = \min_{j \ne i}^{j 
\circ\hskip-1.25pt-\hskip-1.7pt\raisebox{-0.8pt}{\scalebox{0.75}{\mbox{\rotatebox{45}{$\diamond$}}}} 
\alpha} \big| \eta_{j \to \alpha}^{(k)} \big| \cdot \prod_{j \ne i}^{j 
\circ\hskip-1.25pt-\hskip-1.7pt\raisebox{-0.8pt}{\scalebox{0.75}{\mbox{\rotatebox{45}{$\diamond$}}}} 
\alpha} \operatorname{sign} \eta_{j \to \alpha}^{(k)} \NN
  \end{align*} At the beginning of the decoding there are no messages to 
bits, \IE, $\mu_{\alpha \to i}^{(-1 / 2)} \equiv 0$.

Let us define the noise vector $\VEC{\xi} = (\xi_1, \xi_2, \dots, 
\xi_N)$ by $\xi_i = 1 - \sigma_i x_i$, $i = 1, 2, \dots, N$. For 
simplicity, the channel is assumed to be symmetric, $P(-x|+) = P(x|-)$, 
then the decoding error probability (and error causing noise 
configurations) is independent of the codeword $\VEC{\sigma}$ being 
sent.

Consider the error correcting code, the transmission channel, and the 
decoding algorithm (including the [maximal] number of iterations) being 
fixed. The channel noise space $\REAL^N$ is then divided into two sets: 
$\REAL^N \backslash {\cal E}$ and ${\cal E}$, noise realizations that 
are decoded successfully and the ones that result in the decoding error.

\begin{definition} An \emph{instanton} is a noise configuration 
$\VEC{\xi}$ such that: (1) it belongs to the closure of the set ${\cal 
E}$; (2) there is a sufficiently small vicinity of $\VEC{\xi}$ such that 
inside it there are no points of ${\cal E}$ with $P$ larger than 
$P(\VEC{\xi})$. \end{definition}

The instantons are defined as the positions of local maxima of the noise 
distribution density $P(\VEC{\xi}) = \prod_{i = 1}^N P(1 - \xi_i | +)$ 
over the set of error causing noise configurations ${\cal E}$. In the 
limit of high SNR the probability of the decoding error somewhere in the 
information block, Frame-Error Rate (FER), is controlled by the 
instanton with maximal $P(\VEC{\xi})$ and its vicinity. (In order to 
describe the FER {\it vs.} SNR dependence in the moderate SNR region one 
may need to collect the contribution from several instantons.) Such a 
definition of instanton is a paraphrasing of ``source of trouble'', and 
is practically useless without a method to locate it. There were 
numerous methods devised to find/enumerate trapping/stopping/absorbing 
sets and pseudo-codewords of LDPC codes, \EG, \cite{2009_VCNP, 
2010_IEEETIT_DZAWN, 2012_IEEETIT_KB, 2014_IEICE_GMJL, 2016_IEEETIT_LD}.

The min-sum decoding is the high SNR limit of the sum-product algorithm. 
In addition, if $P(1 - \xi|+) = \exp \bigl( -\beta({\rm SNR}) \cdot 
F(\xi) \bigr) / Z({\rm SNR})$ for some increasing function $\beta({\rm 
SNR})$, then the decoding input has the form $h(\xi) = \beta \cdot \big( 
F(2 - \xi) - F(\xi) \big) / 2$. (This includes Additive White Gaussian 
Noise (AWGN) channel with $F(\xi) = \xi^2 / 2$ and $h = \beta (1 - 
\xi)$.) As the min-sum decoding is scalable (\IE, the result of the 
decoding stays the same if the decoding input vector $\VEC{h}$ is 
multiplied by a positive number), the set ${\cal E}$ is independent of 
${\rm SNR}$, so are the instantons.

\section{Cycling of iterations}

One could imagine several possibilities how the iterative decoder could 
fail:
 \begin{itemize}
  \item[{\sf R}:] The iterative decoding {\em was converging to the 
right solution}, but it {\em didn't succeed} during the allowed number 
of iterations.
  \item[{\sf W}:] The iterative decoding {\em converged but to a wrong 
place}. After the convergence the decoding output is a {\em codeword}, 
just not the one that was sent.
  \item[{\sf P}:] The iterative decoding {\em converged but to a wrong 
place}. After the convergence the decoding output is {\em not even a 
codeword}.
  \item[{\sf C}:] The iterative decoding {\em is not going to converge} 
no matter how many iterations you can afford.
 \end{itemize}

The situation {\sf R}{\small [ight]} can be corrected by adding more 
iterations. It is highly possible that in situation {\sf W}{\small 
[rong]} even maximum likelihood decoding would make an error, and the 
probability of such a situation [in the presence of error floor] is very 
small, thus the error because of possibilities {\sf P}{\small 
[seudo-codeword]} or {\sf C}{\small [ycling]} is much more probable.

Following the so-called Bethe free energy variational approach 
\cite{2005_IEEETIT_YFW}, belief propagation can be understood as a set 
of equations for beliefs solving a constrained minimization problem. On 
the other hand, a more traditional approach is to interpret belief 
propagation in terms of an iterative procedure --- so-called belief 
propagation iterative algorithm \cite{63Gal,88Pea,1999_IEEETIT_M}. Being 
identical on a tree (as then belief propagation equations are solved 
explicitly by iterations from leaves to the tree center) the two 
approaches are however distinct for a graphical problem with loops. In 
case of their convergence, belief propagation algorithms find a minimum 
of the Bethe free energy \cite{2005_IEEETIT_YFW,UAI2002_TJ,2004_NC_H}, 
however in a general case convergence of the standard iterative belief 
propagation is not guaranteed.

Experiments with the Tanner's $[155,64,20]$ code \cite{tannercode} 
showed the following: The instanton for linear programming decoding 
\cite{2005_IEEETIT_FWK}, that is minimizing a certain part of the Bethe 
free energy and is not iterative in nature, for AWGN channel has the 
effective distance close to $16.4$ 
\cite{2003_KV,2008_IEEETIT_CS,2012_IEEETC_LB}. At the same time the 
noise configuration $\VEC{\xi}$ with effective distance or weight 
$w(\VEC{\xi}) = \Vert \VEC{\xi} \Vert_2^2 \approx 12.45$ which 
withstands $410$ iterations was found \cite{ISIT2006_SC}. There is a 
strong indication that in the close vicinity of this noise configuration 
there are ones that withstand arbitrarily large number of iterations.

If the decoder provides errors mostly due to situation {\sf P}, then it 
converges in most occasions. The fixed point of iterative decoding is 
the minimum of Bethe free energy. Thus, the iterative decoder should 
work not worse than linear programming decoder, as the latter neglects a 
certain part of Bethe free energy. That contradicts to what was observed 
experimentally for the Tanner's $[155,64,20]$ code: $12.45 < 16.4$.

In contrast with the decoding algorithms which are static ({\it e.g.}, 
linear programming decoding), in the case of iterative decoding the 
instantons could be inherently dynamic, and in order to find them the 
dynamics of iterations [in full details] should necessarily be 
considered.

As an example of cycling of iterations, consider a simple code with $4$ 
bits and $5$ parity checks:
 \BEA
   & {\displaystyle {\hat H} = \left( \begin{array}{cccc}
       1 & 1 & 0 & 0 \\
       0 & 1 & 1 & 0 \\
       0 & 0 & 1 & 1 \\
       1 & 0 & 0 & 1 \\
       1 & 1 & 1 & 1 \\
     \end{array} \right) . }
       \hskip30pt
     \raisebox{-30pt}{\includegraphics[width=68pt]{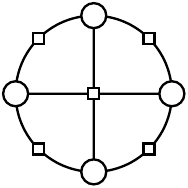}} & \NN
  \EEA The parity checks are obviously redundant, and the code has $2$ 
codewords: $(+, +, +, +)$ and $(-, -, -, -)$. This $[4, 1, 4]$ 
repetition code was one of the examples in \cite{2007_IEEETIT_KS}. As 
the first $4$ parity checks have connectivity $2$, the only linear 
programming pseudo-codewords are the codewords. Because of the checks 
with connectivity $2$ all the bits should have the same 
values.\footnote{It is possible for different copies of a bit to be 
assigned differently on an $n$-cover of the code, $n > 1$, but the 
number of ``$+$''s and ``$-$''s will be the same for all $4$ original 
bits.} In \cite{2007_IEEETIT_KS} more general bit assignments on the 
code's computational tree are considered.

The lowest instanton that survives infinite number of iterations is 
$\VEC{\xi} = (10, 6, 4, 4) / 7$ with the weight $w(\VEC{\xi}) = (10^2 + 
6^2 + 4^2 + 4^2) / 7^2 = 168 / 7^2 = 24 / 7 < 4$ (the numeration of bits 
goes along the $8$-cycle containing the checks with connectivity $2$).

 \begin{figure}[!t]
 \BEA
   & {\displaystyle \begin{array}{r||r|r|r|r}
     \phantom{|_{0_{0_0}}}k & m_1^{(k)} & m_2^{(k)} & m_3^{(k)} & 
m_4^{(k)} \\
     \hline
     \!\!\!\!\!\phantom{{}^{1^1}}0 & -3  &  1  &  3  &  3 \\
     1 &  2  & -2  &  6  &  2 \\
     2 &  4  &  8  & -2  &  6 \\
     3 &  8  &  4  & 12  & -2 \\
     4 & -2  & 12  &  4  & 20 \\
     5 & 30  & -2  & 14  &  4 \\
     6 &  4  & 38  & -2  & 18 \\
     7 & 20  &  4  & 48  & -2 \\
     8 & -2  & 24  &  4  & 56 \\
     \cdots & \cdots & \cdots & \cdots & \cdots \\
\hskip-5pt     4 n &       -2 &     12 n &         4 & 36 n - 16 \\
\hskip-5pt 4 n + 1 & \phantom{1}36 n - 6 &       -2 &  12 n + 2 &         4 \\
\hskip-5pt 4 n + 2 &        4 & \phantom{1}36 n + 2 &        -2 &  12 n + 6 \\
\hskip-5pt 4 n + 3 & 12 n + 8 &        4 & 36 n + 12 &        -2
     \end{array} } & \NN
 \EEA
 \caption{Decoding dynamics on the instanton $\VEC{\xi} = (10, 6, 4, 4) 
/ 7$. The vector $\VEC{h}$ is proportional to $(-3, 1, 3, 3)$, and (as 
the decoding is scalable) the latter was used as $\VEC{h}$ to form the 
table. The general formula at the end starts to be applicable from $n 
\ge 1$, while at iteration $k = 2$ it is not valid yet.}
  \label{fig:cycling}
 \end{figure}

The cycling dynamics of iterations is shown at Fig.~\ref{fig:cycling}. 
The decoding output (and the messages $\mbox{bits} \leftrightarrow 
\mbox{checks}$) is not exactly periodic with the iteration number. If 
one considers one iteration of the decoder as a mapping in the space of 
messages $\VEC{\eta}$, then the instantons are not necessarily periodic 
orbits (\IE, exact cycles) of the mapping.

\section{Instantons for AWGN channel} \label{col_AWGN}

The method of \cite{2008_IEEETIT_CS} (which is improved in 
\cite{2011_ISIT_CS}) to find low weight linear programming decoding 
pseudo-codewords is quite effective. What makes it possible is an easy 
way to convert the output of the LP decoding (a pseudo-codeword) to the 
minimal norm noise with the same [or lower weight] decoding output. Here 
it is shown how to search for low weight instantons for iterative 
decoding in a similar fashion. (A modified, unfruitful but not 
uninteresting, version of Sec.~\ref{col_AWGN} is discussed in Appendix.) 
In contrast to Sec.~\ref{INST_ARRAY}, all the $n_{\rm iter}$ iterations 
in the min-sum decoding are executed, regardless of what could be the 
decoding output in the middle of the decoding process.

The result of the iterative decoding at a certain bit after $n_{\rm 
iter}$ iterations coincides with the result of the iterative decoding on 
a unwrapped Tanner graph of the code (computational tree) with $n_{\rm 
iter}$ generations with this bit at its center; and the performance of 
the code is determined by the effective weights of the codewords on 
computational tree \cite{Wiberg, 2007_IEEETIT_KS}.

\begin{definition} In the case of min-sum decoding, the decoding output 
$m_i^{(k)}$ [at any bit $i$ and at each iteration $k$] is always a 
linear combination of the decoder inputs $h_j$, $1 \le j \le N$, with 
integer coefficients: $m_i^{(k)} = \sum_{j = 1}^N n_{i, 
j}^{(k)}(\VEC{h}) \, h_j = \VEC{n}_i^{(k)}(\VEC{h}) \cdot \VEC{h}$. A 
\emph{colored structure} associated with the input $\VEC{h}$ and the 
output $m_i^{(k)}$ is the $N$-dimensional integer vector 
$\VEC{n}_i^{(k)}(\VEC{h})$.\footnote{In \cite{2005_PRL_SCCV} a 
\emph{colored structure} is the subset of bits of the computational tree 
${\rm CT}_i$ that make a non-zero contrubution to $m_i^{(k)}$.} 
\end{definition}

The colored structure can be computed using a modified version of the 
min-sum iterative decoding: \BEA \!\!\!\! & {\displaystyle\mbox{decoding 
  output:}} &
    \VEC{n}_i^{(k)} = \VEC{e}_i + \sum_{\alpha}^{i
\circ\hskip-1.25pt-\hskip-1.7pt\raisebox{-0.8pt}{\scalebox{0.75}{\mbox{\rotatebox{45}{$\diamond$}}}} 
\alpha} \VEC{\mu}_{\alpha \to i}^{(k - 1/2)} \NN \\
    \!\!\!\! & {\displaystyle\begin{array}{c} \raisebox{-1pt}{\mbox{bits 
$\to$ checks}} \\ \raisebox{1pt}{\mbox{messages}} \end{array}\!\!\!\!:} 
& \VEC{\eta}_{i \to \alpha}^{(k)} = \VEC{e}_i + \sum_{\beta \ne 
\alpha}^{i 
\circ\hskip-1.25pt-\hskip-1.7pt\raisebox{-0.8pt}{\scalebox{0.75}{\mbox{\rotatebox{45}{$\diamond$}}}} 
\beta} \VEC{\mu}_{\beta \to i}^{(k - 1/2)} \NN \\
    \!\!\!\! & {\displaystyle\begin{array}{c} 
\raisebox{-1pt}{\mbox{checks $\to$ bits}} \\ 
\raisebox{1pt}{\mbox{messages}} \end{array}\!\!\!\!:} & 
\VEC{\mu}_{\alpha \to i}^{(k + 1 / 2)} = \VEC{\eta}_{j \to \alpha}^{(k)} 
\prod_{l \ne i, j}^{l 
\circ\hskip-1.25pt-\hskip-1.7pt\raisebox{-0.8pt}{\scalebox{0.75}{\mbox{\rotatebox{45}{$\diamond$}}}} 
\alpha} \operatorname{sign} \bigl( \VEC{h} \cdot \VEC{\eta}_{l \to 
\alpha}^{(k)} \bigr) \NN \\
  & \mathrlap{\mbox{~~~~~~~where } j 
\circ\hskip-2.7pt-\hskip-2.3pt\raisebox{-0.9pt}{\mbox{\scalebox{0.95}{\rotatebox{45}{$\diamond$}}}} 
\alpha \mbox{ and } \bigl| \VEC{h} \cdot \VEC{\eta}_{j \to \alpha}^{(k)} 
\bigr| \mbox{ is minimal}} \NN \EEA
 At the beginning of the decoding there are no messages to bits, 
$\VEC{\mu}_{\alpha \to i}^{(-1 / 2)} \equiv \VEC{0}$. All the messages 
are vectors of length $N$. They contain a detailed enough information 
about how strongly each bit affects the min-sum decoding output at any 
other bit (\EG, $n^{(k)}_{i,j}(\VEC{h})$ is how $h_j$ affects the output 
at bit $i$ after $k$ iterations). By decomposing the aggregated messages 
$m_i^{(k)}$ into components $n_{i,j}^{(k)}$ (\IE, the colored 
structure), such a decoding extracts enough data about the 
pseudo-codeword on the computational tree that causes a decoding error.

Here is how the minimal $L^2$-norm noise vector $\VEC{\xi}$, an 
instanton for AWGN channel, causing an error at bit $i$ is obtained from 
the colored structure \cite{Wiberg}: We consider
 \begin{align*}
   \mathcal{F}_i(\VEC{\xi}) = \frac12 \sum_{j = 1}^N \xi_j^2
+ \lambda \sum_{j = 1}^N (1 - \xi_j) \, n_{i,j}^{(n_{\rm iter})}(\VEC{1} 
- \VEC{\xi})
 \end{align*} Here $1 - \xi_j = h_j$ is the decoder input, and $\lambda$ 
is the Langrange multiplier for the condition $\VEC{h} \cdot 
\VEC{n}_i^{(n_{\rm iter})} = 0$, \IE, the output $m^{(n_{\rm iter})}_i$ 
at the $i^{\rm th}$ bit is completely undecided. Minimizing 
$\mathcal{F}_i$ with respect to $\VEC{\xi}$, we get $\VEC{\xi} = \lambda 
\VEC{n}_i^{(n_{\rm iter})}$. Enforcing the condition $(\VEC{1} - 
\VEC{\xi}) \cdot \VEC{n}_i^{(n_{\rm iter})} = 0$, we get $\lambda = 
\sum_{j = 1}^N n_{i, j}^{(n_{\rm iter})} / \sum_{j = 1}^N \bigl( n_{i, 
j}^{(n_{\rm iter})} \bigr)^2$.

In the case of several colored structures being compatible with the 
decoder input $\VEC{h}$ (this happens if, while computing check to bit 
message, the quantity $\bigl| \VEC{h} \cdot \VEC{\eta}_{j \to 
\alpha}^{(k)} \bigr|$ has the same value for different $j$) the function 
$\mathcal{F}_i$ is modified to
 \begin{align*}
   \mathcal{F}_i(\VEC{\xi}) = \frac12 \sum_{j = 1}^N \xi_j^2 
+ \sum_{c = 1}^C \lambda_c \sum_{j = 1}^N (1 - \xi_j) n_{i,j;\, 
c}^{(n_{\rm iter})}(\VEC{1} - \VEC{\xi}) \end{align*} Here $C$ is the 
number of colored structures competing with each other. Minimizing 
$\mathcal{F}_i$ with respect to $\VEC{\xi}$, we get $\VEC{\xi} = \sum_{c 
= 1}^C \lambda_c \VEC{n}_{i;\, c}^{(n_{\rm iter})}$. Enforcing the 
condition $(\VEC{1} - \VEC{\xi}) \cdot \VEC{n}_{i;\, c}^{(n_{\rm iter})} 
= 0$ for all the $C$ colored structures, we get the system of $C$ linear 
equations $\sum_{d = 1}^C \mathcal{A}_{cd} \lambda_d = \mathcal{B}_c$ 
for $C$ unknowns $\lambda_1$, $\lambda_2$, ..., $\lambda_C$; where
 \begin{align} \mathcal{A}_{cd} = \sum_{j = 1}^N n_{i,j;\,
c}^{(n_{\rm iter})} n_{i,j;\, d}^{(n_{\rm iter})}, \qquad \mathcal{B}_c = 
\sum_{j = 1}^N n_{i,j;\, c}^{(n_{\rm iter})} \label{A_B}
 \end{align} While \cite{2005_PRL_SCCV} generalizes the formula of 
Wiberg \cite{Wiberg} for the case of some bits on a computational tree 
contributing negatively; the formulas $\mathcal{A}_{cd} = \sum_{j = 1}^N 
n_{j;\, c} n_{j;\, d}$, $\mathcal{B}_c = \sum_{j = 1}^N n_{j;\, c}$, 
$\sum_{d = 1}^C \mathcal{A}_{cd} \lambda_d = \mathcal{B}_c$, $\VEC{\xi} 
= \sum_{c = 1}^C \lambda_c \VEC{n}_{c}$ (for brevity the lower index $i$ 
and the upper index $n_{\rm iter}$ are omitted here) generalize it for 
the case when which part of the computational tree contributes to the 
output at the central bit is ambiguous.

An example of two colored structures competing with each other is the 
``instanton (b)'' from \cite{2005_PRL_SCCV}. With the bits numeration 
used in \cite{2005_PRL_SCCV}, it corresponds to a noise configuration 
producing a decoding error at bit $2$. The two colored structures (here 
named ``pink'' and ``yellow'' due to bits coloring in \cite[Figs.~1(b) 
and S6]{2005_PRL_SCCV}) differ in bits $0$ and $77$. Here is the list of 
bits participating in the colored structures:
 \begin{align*}
  \begin{array}{r|ccccccccccccc}
    j & {\it 0} & {\bf 2} & 4 & \!\!11\!\! & \!\!12\!\! & \!\!17\!\! & 
\!\!36\!\! & \!\!56\!\! & \!\!{\it 77}\!\! & \!\!89\!\! & 
\!\!\!117\!\!\! & \!\!\!139\!\!\! & \!\!\!141\!\!\! \\
  \hline
    \!\!n^{(4)}_{2,j; \,{\rm pink}}\!\! & {\it 6} & {\bf 5} & 1 & 1 & 5 
& 1 & 1 & 6 & {\it 4} & 5 & 4 & 4 & 3 \\
    \!\!n^{(4)}_{2,j; \,{\rm yellow}}\!\! & {\it 5} & {\bf 5} & 1 & 1 & 
5 & 1 & 1 & 6 & {\it 3} & 5 & 4 & 4 & 3
   \end{array} \NN
 \end{align*} The $2 \times 2$ system for Lagrange multipliers looks 
like \begin{align*}
   \left[ \begin{array}{rr}
     \!208\! & \!198\! \\
     \!198\! & \!190\!
   \end{array} \right] \left[ \begin{array}{c}
     \!\lambda_{\rm pink}\! \\ \!\lambda_{\rm yellow}\!
   \end{array} \right] =
   \left[ \begin{array}{c} \!46\! \\ \!44\! \end{array} \right] \NN
 \end{align*} with $\lambda_{\rm pink} = 7 / 79$ and $\lambda_{\rm 
yellow} = 11 / 79$ being the solution. The corresponding instanton is 
$\VEC{\xi}_{(b)} = \lambda_{\rm pink} \VEC{n}^{(4)}_{2; \,{\rm pink}} + 
\lambda_{\rm yellow} \VEC{n}^{(4)}_{2; \,{\rm yellow}}$. It has the 
effective weight $\Vert \VEC{\xi}_{(b)} \Vert_2^2 = 806 / 79 \approx 
10.2025$.

The algorithm shown in Fig.~\ref{fig:procedure_AWGN} is an adaptation of 
the LP decoding pseudo-codewords search method of \cite{2008_IEEETIT_CS} 
to the case of min-sum iterative decoder. The necessity of cautious 
update of $\VEC{\xi}$, see Fig.~\ref{fig:update}, arises from the fact 
that the noise configuration $\VEC{\xi}_{\rm temp}$ calculated in {\sf 
L5} could produce a different colored structure in decoding, see, \EG, 
\cite[Fig.~S6]{2005_PRL_SCCV}. Whenever the parameter $t$ in the update 
is too small (\IE, we can not shift $\VEC{\xi}$ further towards 
$\VEC{\xi}_{\rm temp}$ and still have a decoding error), the colored 
structure computed in {\sf L4} and the noise $\VEC{\xi}_{\rm temp}$ are 
not compatible. This suggests that the current instanton candidate 
$\VEC{\xi}$ corresponds to several colored structures, thus the line 
{\sf L8}.

While generating Figs.~\ref{fig:progress_AWGN} and 
\ref{fig:spectra_AWGN}, the noise $\VEC{\xi}$ in {\sf L1} was produced 
by AWGN channel with ${\rm SNR} = 1$ until the decoding error was 
detected. The ``small noise'' in {\sf L8} was AWGN with standard 
deviation $10^{-8}$. Removal of redundant colored structures in {\sf L9} 
was done by rank-revealing QR factorization.

 \begin{figure}[!t]
 \centering
 \begin{tabular}{r|l}
   {\sf L1} & $t \leftarrow 1$ \\
   {\sf L2} & while ($t \mkern1mu \VEC{\xi}_{\rm temp} + (1 - t) 
\mkern1mu \VEC{\xi}$ does not cause error) \\
   {\sf L3} & \hskip18pt $t \leftarrow t / 2$ \\
   {\sf L4} & $\VEC{\xi} \leftarrow t \mkern1mu \VEC{\xi}_{\rm temp} + 
(1 - t) \mkern1mu \VEC{\xi}$ \\
 \end{tabular}
 \caption{Updating the noise vector $\VEC{\xi}$ from $\VEC{\xi}_{\rm 
temp}$.}
 \label{fig:update}
 \end{figure}

 \begin{figure}[!t]
 \centering
 \begin{tabular}{r|l}
   {\sf L1} & start with some error causing noise vector $\VEC{\xi}$ \\
   {\sf L2} & $\VEC{h} \leftarrow \VEC{1} - \VEC{\xi}$ \\
   {\sf L3} & $i \leftarrow \argmin_i \smash{m_i^{(n_{\rm iter})}}$ \\
   {\sf L4} & compute the color structure 
$\smash{\VEC{n}_{i}^{(n_{\rm iter})}}(\VEC{h})$ \\
   {\sf L5} & $\VEC{\xi}_{\rm temp} \leftarrow 
\smash{\VEC{n}_{i}^{(n_{\rm iter})}} \sum_{j = 1}^N \smash{n_{i, 
j}^{(n_{\rm iter})}} / \sum_{j = 1}^N \smash{\bigl( n_{i, j}^{(n_{\rm 
iter})} \bigr)^2}$ \\
   {\sf L6} & update $\VEC{\xi}$ from $\VEC{\xi}_{\rm temp}$, see 
Fig.~\ref{fig:update} \\
   {\sf L7} & if ($t$ is very small (\EG, $t < 10^{-8}$)) then \\
   {\sf L8} & \hskip18pt create a list of colored structures 
$\smash{\VEC{n}_{i;\, c}^{(n_{\rm iter})}}$ \\
     & \hskip36pt by computing $\smash{\VEC{n}_{i}^{(n_{\rm 
iter})}}(\VEC{1} - \VEC{\xi} + \text{small noise})$ \\
     & \hskip36pt for several realizations of small noise \\
   {\sf L9} & \hskip18pt remove linearly dependent colored structures 
\\
   {\sf L10} & \hskip18pt form $\mathcal{A}_{cd}$ and $\mathcal{B}_c$ as 
in eq.~(\ref{A_B}) \\
   {\sf L11} & \hskip18pt $\VEC{\xi}_{\rm temp} \leftarrow \sum_{c} 
\lambda_c \smash{\VEC{n}_{i;\, c}^{(n_{\rm iter})}}$, where $\sum_d 
\mathcal{A}_{cd} \lambda_d = \mathcal{B}_c$ \\
   {\sf L12} & \hskip18pt update $\VEC{\xi}$ from $\VEC{\xi}_{\rm 
temp}$, see Fig.~\ref{fig:update} \\
   {\sf L13} & go to {\sf L2} or exit
 \end{tabular}
 \caption{Instantons for AWGN channel search algorithm.}
 \label{fig:procedure_AWGN}
 \end{figure}

The characteristic feature of Fig.~\ref{fig:progress_AWGN} is sudden 
drops of the effective weight $w = \Vert \VEC{\xi} \Vert_2^2$. They 
happen when the noise configuration is updated in Fig.~\ref{fig:update} 
with the value of $t$ being not much smaller than $1$. Such drops due to 
updates in both {\sf L6} and {\sf L12} were observed.

The Fig.~\ref{fig:spectra_AWGN} could be compared to 
\cite[Fig.~4]{ISIT2006_SC}. The weight distribution of the resulting 
error causing noise configurations is worse for $n_{\rm iter} = 4$ and 
better for $n_{\rm iter} = 10$, if compared to \cite{ISIT2006_SC}, where 
the distributions produced by instanton-amoeba scheme for $n_{\rm iter} 
= 4$ and $n_{\rm iter} = 8$ are shown (while $n_{\rm iter} = 20$ was 
infeasible for amoeba). The algorithm shown in 
Fig.~\ref{fig:procedure_AWGN} takes considerably less time to converge, 
although no scrutinous quantitative comparison was done.

Experiments did show that for very large $n_{\rm iter}$ the procedure 
could get stuck for considerable number of steps, due to myriads of not 
too low weight noise configurations corresponding to multiple colored 
structures.

 \begin{figure}[!t] \centering 
 \includegraphics[width=\columnwidth]{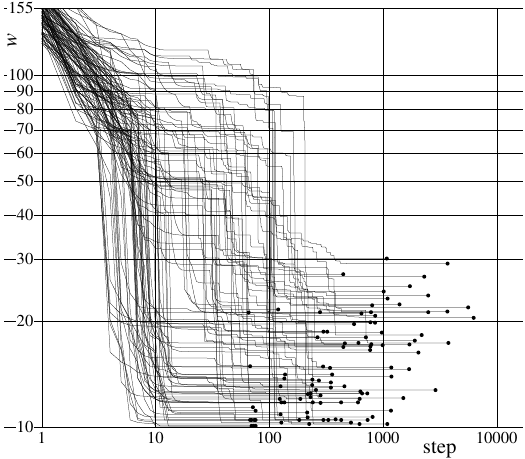} 
 \caption{The effective weight $w = \Vert \VEC{\xi} \Vert_2^2$ for
$n_{\rm iter} = 4$ decoder and the Tanner's $[155, 64, 20]$ code {\it 
vs.} the number {\sf L2} line in Fig.~\ref{fig:procedure_AWGN} visits, 
\IE, the number of algorithm steps, $100$ realizations are shown.}
 \label{fig:progress_AWGN}
 \end{figure}

 \begin{figure}[!t]
 \centering
 \includegraphics[width=\columnwidth]{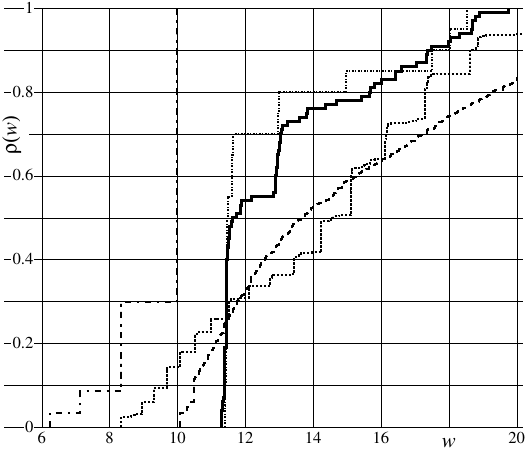}
 \caption{The probability/frequency of occurrence, $\rho(w)$, of the 
produced by the search algorithm in Fig.~\ref{fig:procedure_AWGN} 
instantons to have the weight $w$ or smaller, $n_{\rm iter} = 2$, $3$, 
$4$, $10$, and $20$ (dash-dotted, dotted, dashed, solid, and fuzzy line, 
and $1000$, $1000$, $1000$, $100$, and $20$ realizations, 
respectively).}
 \label{fig:spectra_AWGN}
 \end{figure}

\section{Instantons array} \label{INST_ARRAY}

In this section the iterative decoder after each iteration checks the 
output $\VEC{\sigma} = \operatorname{sign} \VEC{m}$ for being a valid 
codeword, \IE, ${\hat H} \cdot \big( (1 - \VEC{\sigma}) / 2 \big) = 
\VEC{0}~\mbox{(mod $2$)}$ (and if it is, the iterations stop).

\begin{definition} The noise configuration is said to {\it withstand} 
$n$ iterations if for all $m \le n$ the decoding output after $m$ 
iterations is wrong (that includes the case when the decoding stops 
before $n$ iterations). All such configurations form a set ${\cal E}(n) 
\subseteq \REAL^N$. \end{definition}

\begin{statement} Checking for the output being a codeword at each 
iteration makes the set ${\cal E}$ being a non-increasing function of 
the number of iterations: ${\cal E}(n + 1) \subseteq {\cal E}(n)$ for 
all $n$. \begin{proof} Let $\VEC{\xi} \notin {\cal E}(n)$, \IE, the 
decoding output at $n^{\rm th}$ [or earlier] iteration is the correct 
one. As it is a valid codeword, the decoding stops, and it will be the 
output of the decoding with any $n_{\rm iter} \ge n$ iterations, \EG, 
$\VEC{\xi} \notin {\cal E}(n + 1)$. \end{proof} \end{statement}

In other words, an error of ${\sf W}${\small [rong]} type is 
unrecoverable, as the decoding is stopped; otherwise doing more 
iterations could potentially result in the correct decoding.

 \begin{figure}[!t]
 \centering
 \begin{picture}(252,437)(0,0)
   \put(18,220.5){\includegraphics[width=216pt]{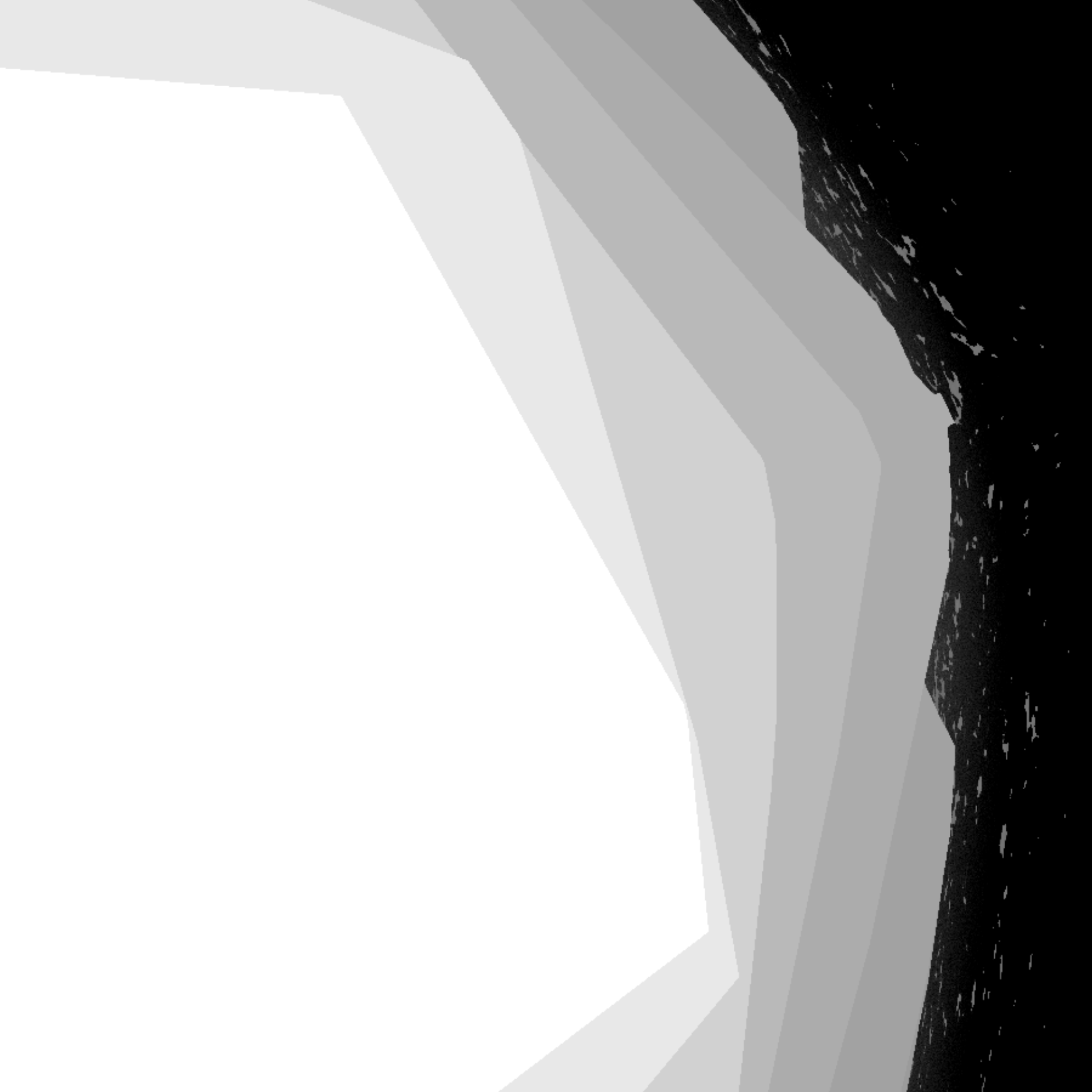}}
   \put(0,220){\includegraphics[width=252pt]{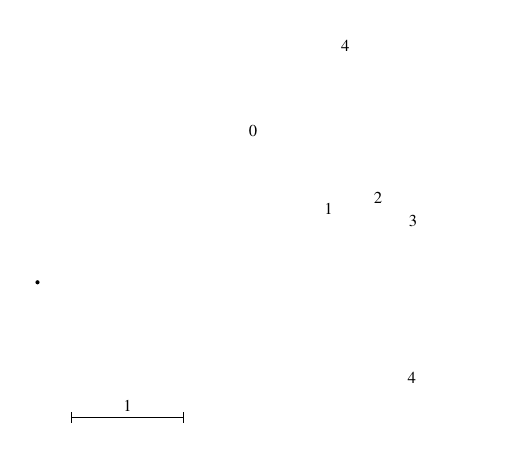}}
   \put(24.5,302){\makebox(0,0)[x]{$\VEC{0}$}}
   \put(200,301){\makebox(0,0)[r]{$\VEC{\xi}_{\sf A}$}}
   \put(176,391){\makebox(0,0)[r]{$\VEC{\xi}_{\sf B}$}} 
   \put(22,432.5){\makebox(0,0)[lt]{\small(a)}}
\put(17,110){\includegraphics[width=108pt]{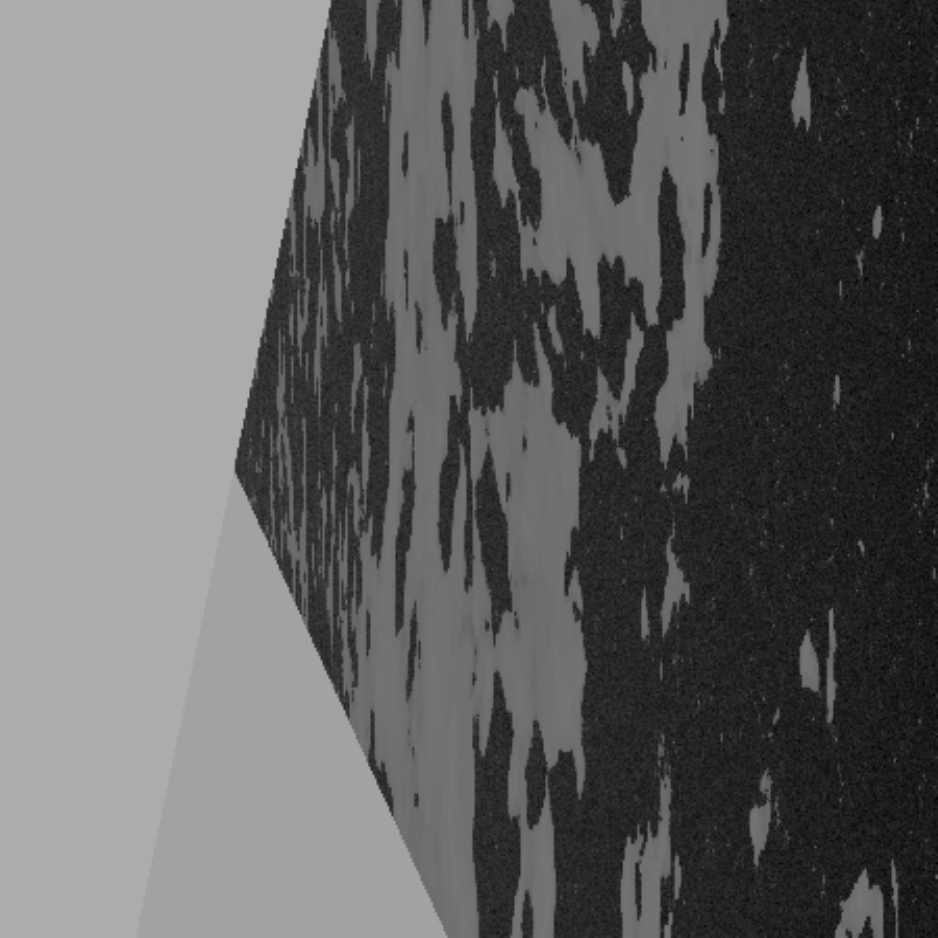}} 
\put(17,110){\includegraphics{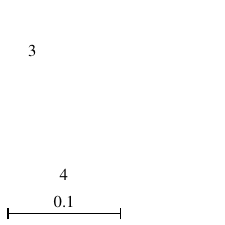}}
   \put(21,214){\makebox(0,0)[lt]{\small(b)}}
\put(127,110){\includegraphics[width=108pt]{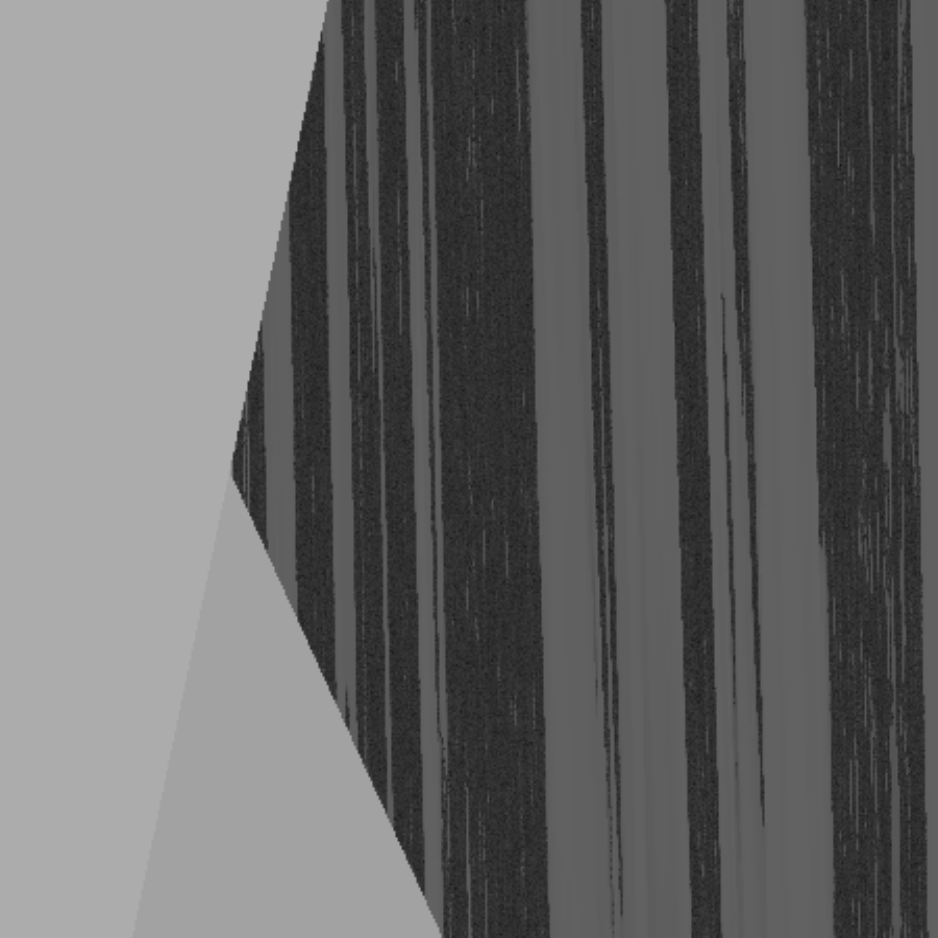}} 
\put(127,110){\includegraphics{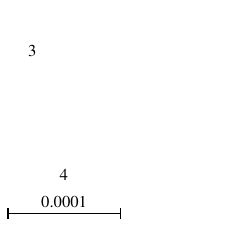}}
   \put(131,214){\makebox(0,0)[lt]{\small(c)}}
\put(17,0){\includegraphics[width=108pt]{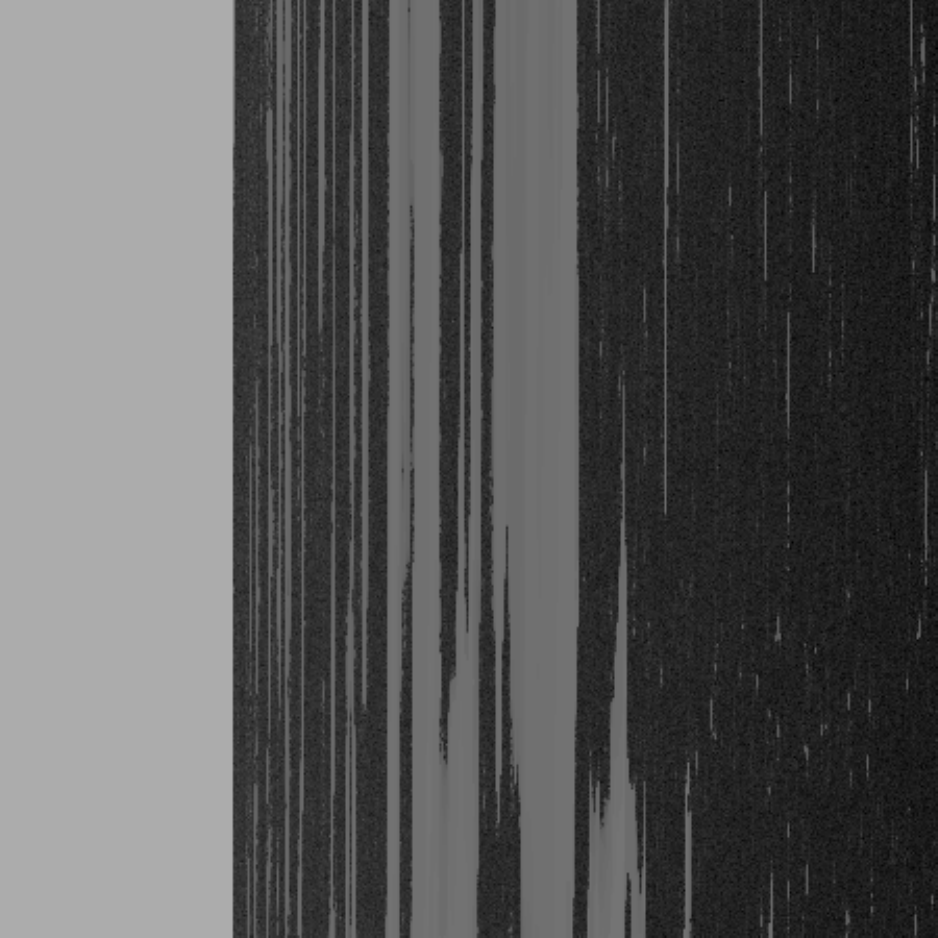}} 
\put(17,0){\includegraphics{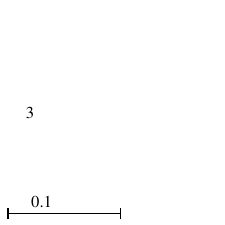}}
   \put(21,104){\makebox(0,0)[lt]{\small(d)}}
\put(127,0){\includegraphics[width=108pt]{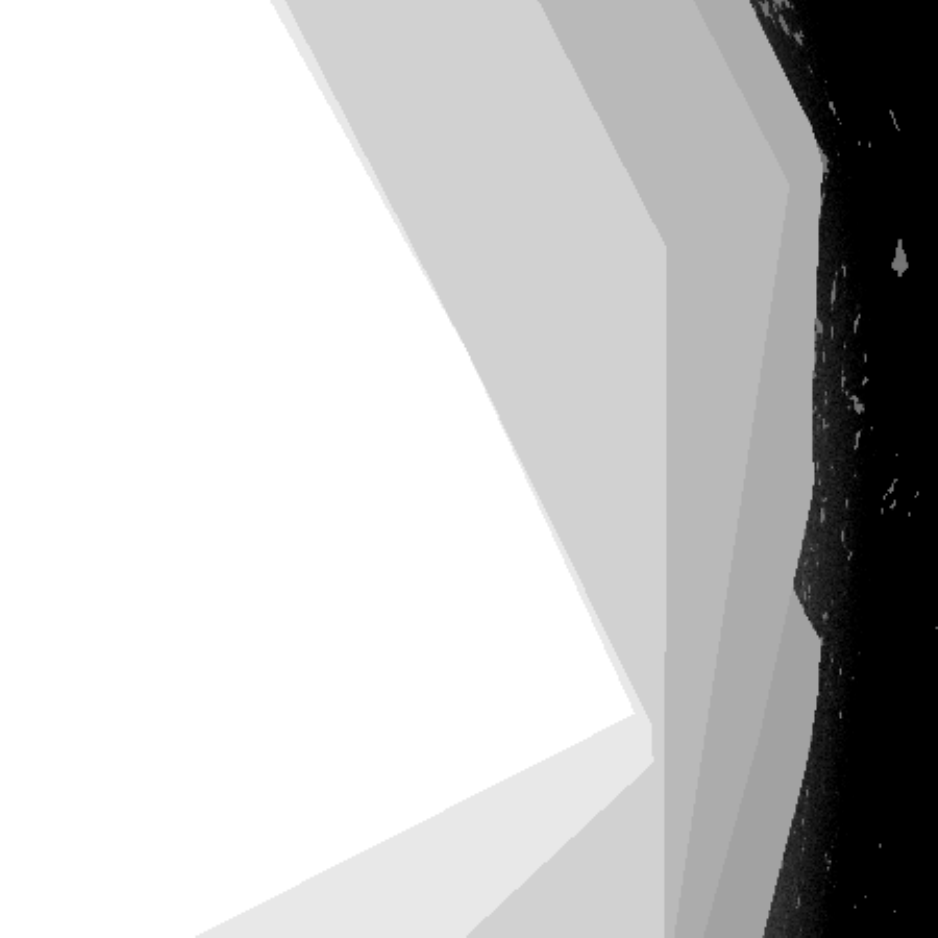}} 
\put(126,-1){\includegraphics{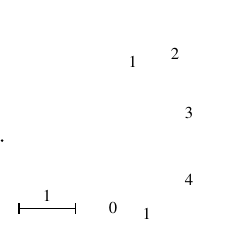}}
   \put(132.5,41){\makebox(0,0)[x]{$\small\VEC{0}$}}
   \put(218,40){\makebox(0,0)[r]{$\small\VEC{\xi}_{\sf A}$}}
   \put(131,104){\makebox(0,0)[lt]{\small(e)}}
 \end{picture}
 \caption{Two-dimensional cuts of the [$155$-dimensional] noise space 
that contain zero noise vector $\VEC{0}$ and the lowest weight instanton 
$\VEC{\xi}_{\sf A}$ for Tanner's $[155, 64, 20]$ code and AWGN channel. 
The line going through $\VEC{0}$ and $\VEC{\xi}_{\sf A}$ is horizontal. 
The plane of the cut is determined by the 3rd point it goes through. In 
panels (a), (b), and (c) it is the instanton $\VEC{\xi}_{\sf B}$; in 
panel (d) it is a random vector; and in panel (e) it is the vector 
$\VEC{t}$ with $t_{60} = t_{122} = t_{130} = t_{131} = t_{136} = 1$ and 
all other components being $0$. The labels $0$, $1$, $2$, $3$, and $4$ 
indicate how many iterations the noise withstands in this area. The tone 
of gray is calculated as $(9 - \log_2 n) / 11$, where $n$ is how many 
iterations the noise configuration withstands, with $0$/$1$ being 
black/white. Tones $10 / 11$ and $1$ correspond to $n = 0$ and correct 
decoding without any iterations (\IE, $\xi_i < 1$ for all $i$).}
 \label{fig:surface}
 \end{figure}

The problem of instanton-amoeba scheme \cite{2005_PRL_SCCV,ISIT2006_SC} 
for large number of iterations is with the rough landscape of the 
function amoeba tries to optimize. The moves amoeba does do assume that 
the landscape is regular (see \cite{1965_CJ_NM,NR}). The problem with 
the application of downhill simplex/amoeba method to finding $n_{\rm 
iter}$ instantons is that amoeba always aims for noise configurations 
that withstand $n_{\rm iter}$ (\IE, many) iterations. The set ${\cal 
E}(n_{\rm iter})$ of such noise configurations [for large $n_{\rm 
iter}$] is very irregular near its boundary (see Fig.~\ref{fig:surface} 
and also \cite{2003_DIMACS_VK}), and amoeba is getting confused and 
uncontrollably reduces its size without any progress.

 \begin{figure}[!t]
 \centering
 \begin{tabular}{r|l}
   {\sf L1} & start with the noise vector $\VEC{\xi} = (1, 1, \dots, 1)$ 
\\
   {\sf L2} & check some (may be empty) list of noise vectors \\
   {\sf L3} & for $k = 0$, $1$, \dots, $n_{\rm iter; max}$ \\
   {\sf L4} & \hskip18pt perturb $\VEC{\xi}(k)$ \\
   {\sf L5} & \hskip18pt check perturbed noise vector \\
   {\sf L6} & go to {\sf L3} or exit
 \end{tabular}
 \caption{Iterative decoding instanton search algorithm.}
 \label{fig:procedure}
 \end{figure}

 \begin{figure}[!t]
 \centering
 \begin{picture}(252,28)(0,0)
\put(14.4,14){\includegraphics[width=3.1in,height=8pt]{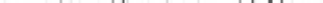}}
\put(14.4,6){\includegraphics[width=3.1in,height=8pt]{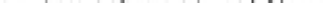}}
 \put(46.8,5){\makebox(0,0)[t]{\scriptsize$23$}}
 \put(56.88,28){\makebox(0,0)[t]{\scriptsize$30$}}
 \put(92.88,28){\makebox(0,0)[t]{\scriptsize$55$}}
 \put(98.64,5){\makebox(0,0)[t]{\scriptsize$59$}}
 \put(100.08,28){\makebox(0,0)[t]{\scriptsize$60$}}
 \put(151.92,5){\makebox(0,0)[t]{\scriptsize$96$}}
 \put(180.72,28){\makebox(0,0)[t]{\scriptsize$116$}}
 \put(189.36,5){\makebox(0,0)[t]{\scriptsize$122$}}
 \put(201.6,28){\makebox(0,0)[t]{\scriptsize$130,131$}}
 \put(209.52,5){\makebox(0,0)[t]{\scriptsize$136$}}
 \put(13.4,19){\makebox(0,0)[r]{$\VEC{\xi}_{\sf A}$}}
 \put(238.6,9){\makebox(0,0)[l]{$\VEC{\xi}_{\sf B}$}}
 \end{picture}
 \caption{The two lowest instantons $\VEC{\xi}_{\sf A}$ and 
$\VEC{\xi}_{\sf B}$. The tone of gray is calculated as $1 - \xi / 2$, 
with $\xi = 2$\;\big/\;$\xi = 0$ being black/white. The $155 \times 93$ 
parity check matrix ${\hat H}$ consists of three blocks: $({\hat R}^1 ~ 
{\hat R}^2 ~ {\hat R}^4 ~ {\hat R}^8 ~ {\hat R}^{16})$, $({\hat R}^5, 
{\hat R}^{10} ~ {\hat R}^{20} ~ {\hat R}^9 ~ {\hat R}^{18})$, $({\hat 
R}^{25} ~ {\hat R}^{19} ~ {\hat R}^7 ~ {\hat R}^{14} ~ {\hat R}^{28})$, 
where ${\hat R}$ is the $31 \times 31$ matrix that cyclically shifts a 
column vector up by one component.}
 \label{fig:two_instantons}
 \end{figure}

 \begin{figure}[!t]
 \centering
 \begin{picture}(252,153)(0,0)
\put(11,146){\includegraphics[width=1.55in,height=7pt]{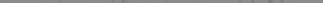}}
\put(127.5,146){\includegraphics[width=1.55in,height=7pt]{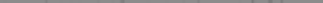}}
\put(11,0){\includegraphics[width=1.55in]{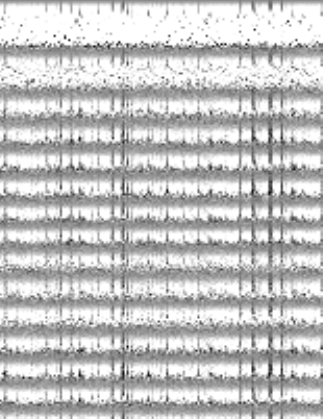}}
\put(127.5,0){\includegraphics[width=1.55in]{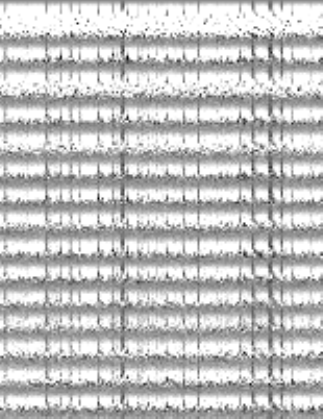}}
 \put(0,0){\makebox(0,0)[lb]{$\VEC{\xi}_{\sf A}$}}
 \put(252,0){\makebox(0,0)[rb]{$\VEC{\xi}_{\sf B}$}}
 \put(10,153){\makebox(0,0)[rt]{\small$\VEC{h}$}}
 \put(241.1,153){\makebox(0,0)[lt]{\small$\VEC{h}$}}
 \end{picture}
 \caption{Iterative decoding output $\VEC{m}$ on the instantons 
$\VEC{\xi}_{\sf A}$ and $\VEC{\xi}_{\sf B}$, $200$ iterations running 
from top to bottom are shown. The tone of gray is calculated as $(1 + m 
/ 10) / 2$, with $0$/$1$ being black/white. Middle gray (tone $1 / 2$) 
corresponds to undecided output $m = 0$. The decoding input $\VEC{h} = 
\VEC{m}^{(0)}$ is shown at the top for comparison of input and output 
magnitudes.}
 \label{fig:decoding}
 \end{figure}

The algorithm shown in Fig.~\ref{fig:procedure} and described below 
overcomes this difficulty and is able to find instantons for large 
$n_{\rm iter}$. The procedure deals with the array of noise 
configurations, $\VEC{\xi}(k)$, $k = 0$, $1$, \dots, $n_{\rm iter;max}$, 
where at any time the noise $\VEC{\xi}(k)$ is the one with the largest 
$P(\VEC{\xi})$ (or the lowest weight $w(\VEC{\xi}) \propto \sum_{i = 
1}^N F(\xi_i)$) from all the withstanding $k$ iterations noise 
configurations that were encountered in the procedure so far. (The 
updates of $\VEC{\xi}(k)$ are done in the line {\sf L5} and (at the 
start) in the line {\sf L2}.)

In the line {\sf L1} of the algorithm the output of the channel is 
completely undecided ($\VEC{h} = (0, 0, \dots, 0)$). This configuration 
obviously withstands $\infty$ iterations, although $P(\VEC{\xi})$ at it 
is quite low. This step makes $\VEC{\xi}(k) = (1, 1, \dots, 1)$ for all 
$k = 0$, $1$, \dots, $n_{\rm iter; max}$.

In the line {\sf L2} the noise configurations that are known from some 
external source ({\rm e.g.}, from previous runs of the procedure or from 
the analysis of trapping sets or pseudo-codewords) may be introduced as 
a starting point for instanton search.

 \begin{figure}[!t]
 \centering
 \includegraphics[width=\columnwidth]{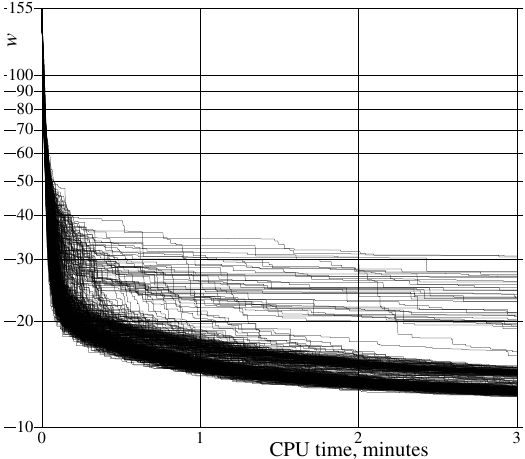}
 \includegraphics[width=\columnwidth]{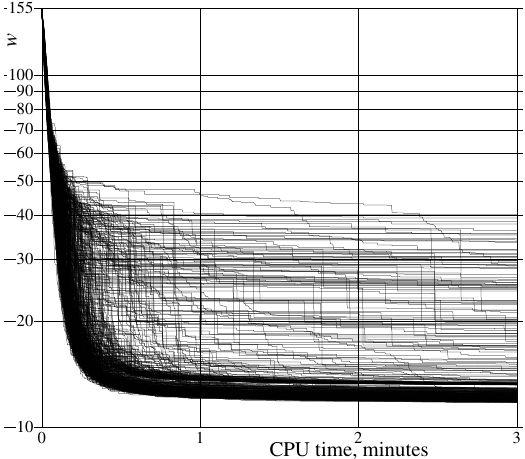}
 \caption{The effective weight $w$ of the withstanding $n_{\rm iter;max} 
= 100$ iterations noise configuration for the Tanner's $[155, 64, 20]$ 
code and AWGN channel {\it vs.} CPU (Intel Xeon X3360, $2.83~{\rm GHz}$) 
time, $500$ realizations are shown. The feedback for the amplitude $a$ 
is {\sf A} (upper panel) and {\sf D} (lower panel).}
 \label{fig:progress}
 \end{figure}

 \begin{figure}[!t]
 \centering
 \includegraphics[width=\columnwidth]{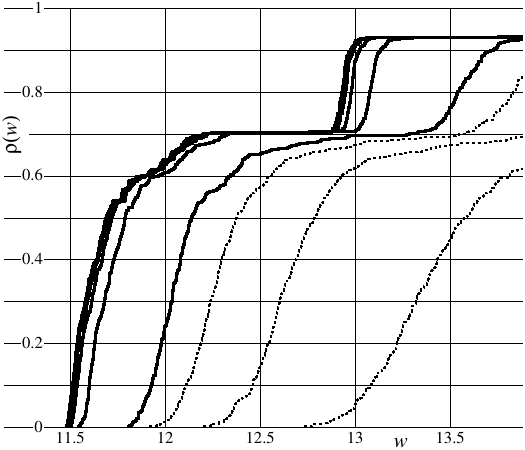}
 \includegraphics[width=\columnwidth]{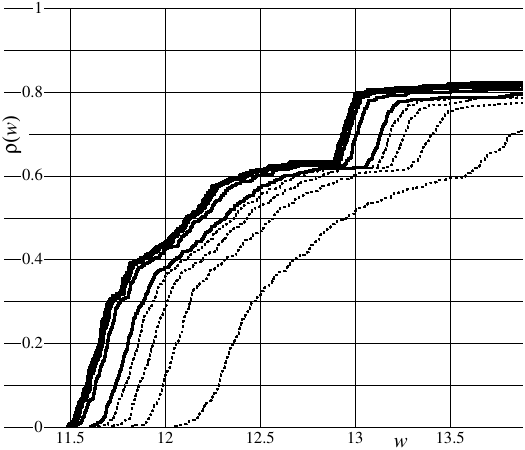}
 \caption{The probability/frequency of occurrence, $\rho(w)$, of the 
withstanding $n_{\rm iter;max} = 100$ iterations noise configurations 
with the weight $w$ or smaller after $1$, $2$, $3$, $4$ (dashed curves) 
and $5$, $10$, \dots, $30$ (solid curves) minutes of CPU time. Code, 
channel, feedback, and CPU are the same as in Fig.~\ref{fig:progress}.}
 \label{fig:spectra}
 \end{figure}

This procedure, applied to Tanner's $[155, 64, 20]$ code and AWGN 
channel, with $n_{\rm iter;max} = 100$, produced an instanton 
$\VEC{\xi}_{\sf A}$ with the lowest weight $w(\VEC{\xi}_{\sf A}) = \Vert 
\VEC{\xi}_{\sf A} \Vert_2^2 < 11.475333$ that causes iterations to cycle 
with the period of length $12$ (see Fig.~\ref{fig:decoding}). The next 
instanton $\VEC{\xi}_{\sf B}$ has the weight $w(\VEC{\xi}_{\sf B}) 
\approx 11.4996$. The differences in weight for configurations that 
withstand $20$ or more iterations are very small. Submitting the array 
$\VEC{\xi}(k)$ as an initial state of the procedure with larger $n_{\rm 
iter;max}$ relatively quickly produces noise configurations with very 
close weight that withstand larger $n_{\rm iter;max}$ iterations.

Below are the details of the procedure used to generate 
Figs.~\ref{fig:progress} and \ref{fig:spectra}.\footnote{The 
perturbation of noise vector in the line {\sf L4} (including the choice 
of the perturbation amplitude), of course, can be done in many different 
ways.} The noise vector $\VEC{\xi}$ is perturbed as $\VEC{\xi} \to c 
\VEC{\xi} + a \VEC{\psi}$, where the components of $\VEC{\psi}$ are 
independent standard normal random variables. The coefficient $c = 
\sqrt{1 - a^2 N / w(\VEC{\xi})} < 1$ makes the expected value ${\sf E} 
\, \Vert c \VEC{\xi} + a \VEC{\psi} \Vert_2^2 = c^2 w(\VEC{\xi}) + a^2 N 
= w(\VEC{\xi})$ not being systematically increased by the addition of $a 
\VEC{\psi}$.

One doesn't want to have the amplitude of the perturbation $a$ being too 
small (or the optimization is slow) or too large (then the perturbed 
noise is rejected often). To accelerate the procedure the amplitude $a$ 
is chosen according to the following negative feedback: Each noise 
configuration $\VEC{\xi}$ has a number $A$ attached to it, and the 
perturbed noise $c \VEC{\xi} + a \VEC{\psi}$ gets the number $2 A$ 
attached, while the number attached to $\VEC{\xi}$ is decreased by a 
factor $0.999$. In the line {\sf L1} the value $A = 0.1$ is attached. 
The amplitude of the perturbation $a$ is chosen as {\sf A}: $a = A$; 
{\sf D}: $0.1 A < a < A$; and {\sf W}: $10^{-14} < a < 0.1$, with 
uniform distribution of $\log a$ in both {\sf D} and {\sf W}. In 
comparison to {\sf A} and {\sf D}, the progress in {\sf W} is slow --- 
the perturbation amplitude $a$ is often too small or too large.

How $w \big( \VEC{\xi}(100) \big)$ goes down with time is shown in 
Fig.~\ref{fig:progress}. It can be seen that sometimes $w \big( 
\VEC{\xi}(100) \big)$ suddenly drops down quite a bit --- it is 
happening when beginning part of the array (but not $\VEC{\xi}(100)$) 
already went lower in weight, and then suddenly a small perturbation 
withstands $100$ iterations, so $\VEC{\xi}(100)$ is updated. Such events 
are what makes the whole procedure work. The progress in the beginning 
part of the array is a lot more regular, and the procedure treasures it 
in hope that it will be converted into the progress at $n_{\rm 
iter;max}$.

The distribution of $w \big( \VEC{\xi}(100) \big)$ is shown in 
Fig.~\ref{fig:spectra}. As it can be seen, the fate of the run is 
determined quite early.

At the very beginning there are few rejections of the perturbed noise 
vectors, and with the feedback {\sf A} [with larger choices of $a$] the 
weight goes down faster than with {\sf D} (see Fig.~\ref{fig:progress}). 
When the weight reaches about $20$, the feedback {\sf D} is more 
effective, probably because eventual smaller than $A$ choices of $a$ 
lead to the perturbations being not rejected more often, which keeps the 
values of $A$ large enough. Eventually {\sf A} is more effective (see 
Fig.~\ref{fig:spectra}), although such a difference between {\sf A} and 
{\sf D} is a bit surprising.

\section{Discussion}

The lowest instanton weight $w$ describes how fast the decoding error 
probability goes down with SNR in the high SNR limit: $\log P({\cal E}) 
\sim -\beta({\rm SNR}) \cdot w$. The value of $w$ quickly saturates with 
the number of iterations $n_{\rm iter}$, and the decrease of $P({\cal 
E})$ with $n_{\rm iter}$ is probably caused by the thinning of the set 
${\cal E}(n_{\rm iter})$ in the vicinity of the instanton. How exactly 
does this happen?

In the vicinity of the instantons $\VEC{\xi}_{\sf A}$ and 
$\VEC{\xi}_{\sf B}$ there are noise configurations that withstand just 
$3$ iterations (see Fig.~\ref{fig:surface}). Is it possible to locate 
these instantons from the analysis of the decoding with small number 
iterations?

In the example from Sec.~II the magnitude of iterative decoder messages 
was growing linearly with the iteration number. Such a linear growth was 
not observed for the instantons $\VEC{\xi}_{\sf A}$ and $\VEC{\xi}_{\sf 
B}$. Should that be expected?

\bibliographystyle{IEEEtran}
\bibliography{refs}

\appendix \label{appendix}

Let ${\rm CT}_i$ be the Computational Tree (CT) with $n_{\rm iter}$ 
generations that starts at the bit $i$; $X$ and ${\cal X}_i$ be the 
vector spaces of noise configurations on the original error correcting 
code and on ${\rm CT}_i$. For any noise configuration $\VEC{\xi} \in X$ 
let us define the noise configuration $\VEC{\Xi} = {\rm 
CT}_i({\VEC{\xi}}) \in {\cal X}_i$ on the ${\rm CT}_i$ by making the 
value of $\VEC{\Xi}$ on any copy of bit $j$ to be equal to $\xi_j$. This 
defines a linear mapping ${\rm CT}_i: X \to {\cal X}_i$.\footnote{The 
following is not going to be used, but is easy to note: The mapping is 
injective for sufficiently large $n_{\rm iter}$ --- when the CT contains 
at least one copy of each bit. Whenever some bit has more than one copy 
on the CT, the mapping is not surjective.}

\begin{definition} A noise configuration $\VEC{\Xi}$ on ${\rm CT}_i$ is 
called \emph{admissible} if the noise values on different copies of the 
same bit are equal, \IE, $\VEC{\Xi} = {\rm CT}_i(\VEC{\xi})$ for some 
noise $\VEC{\xi}$. \end{definition}

Consider some noise configuration $\VEC{\xi}_0$ which causes a decoding 
error after $n_{\rm iter}$ iterations are done, \IE, $\VEC{\xi}_0 \in 
{\cal E}$. The goal is to generate lower weight noise $\VEC{\xi} \in 
{\cal E}$.

For each bit $i$ let us denote the most probable pseudo-codewords 
(codewords on the CT that starts at bit $i$) with correct (``$+$'') and 
incorrect (``$-$'') values at the the central bit by $\VEC{\Sigma}^{(i, 
\pm)}$. (Notice that $\VEC{\Sigma}$, as a vector with components sitting 
on the CT bits, could have different values at different CT copies of 
initially the same bit of the code.) The whole CT structure of 
$\VEC{\Sigma}^{(i, \pm)}$ is not going to be needed. As in 
\cite{Wiberg}, it will be enough to know for each bit $j$ the number 
$M_j^{(i, \pm)}$ of its copies in ${\rm CT}_i$ that have ``$-$'' value 
in $\VEC{\Sigma}^{(i, \pm)}$. The pseudo-codewords $\VEC{\Sigma}^{(i, 
\pm)}$ are then described by $N$-dimensional vectors $\VEC{M}^{(i, 
\pm)}$.

Here is how the vectors $\VEC{M}^{(i, \pm)}$ can be obtained by dynamic 
programming \cite{1957_B}, propagating from the leaves of the tree to 
its center: Proceed with iterative decoding, but instead of standard 
log-likelihood messages send $N$-dimensional vectors containing numbers 
of ``$-$'' copies. Here are the exact formulas for the ``decoding'':
  \begin{align*}
    \!\!\!\! & {\displaystyle\begin{array}{c} \raisebox{-1pt}{\mbox{bits 
$\to$ checks}} \\ \raisebox{1pt}{\mbox{messages}} \end{array}\!\!\!\!:} 
&& \VEC{\eta}_{j \to \alpha}^{(k)}(\pm) = {\TS\frac{1 \mp 1}{2}} 
\VEC{e}_j + \sum_{\beta \ne \alpha}^{j 
\circ\hskip-1.25pt-\hskip-1.7pt\raisebox{-0.8pt}{\scalebox{0.75}{\mbox{\rotatebox{45}{$\diamond$}}}} 
\beta} \VEC{\mu}_{\beta \to j}^{(k - 1/2)}(\pm) \\
    \!\!\!\! & {\displaystyle\begin{array}{c} 
\raisebox{-1pt}{\mbox{checks $\to$ bits}} \\ 
\raisebox{1pt}{\mbox{messages}} \end{array}\!\!\!\!:} && 
\VEC{\mu}_{\alpha \to j}^{(k + 1 / 2)}(\pm) = \sum_{l \ne j}^{l 
\circ\hskip-1.25pt-\hskip-1.7pt\raisebox{-0.8pt}{\scalebox{0.75}{\mbox{\rotatebox{45}{$\diamond$}}}} 
\alpha} \VEC{\eta}_{l \to \alpha}^{(k)}(\sigma_l) \\
  & \mathrlap{\mbox{~~~~where } \prod\limits_{l \ne j}^{l 
\circ\hskip-1.25pt-\hskip-1.7pt\raisebox{-0.8pt}{\scalebox{0.75}{\mbox{\rotatebox{45}{$\diamond$}}}} 
\alpha} \sigma_l = \pm \mbox{~~and~~} \VEC{h} \, \cdot \! \sum_{l \ne 
j}^{l 
\circ\hskip-1.25pt-\hskip-1.7pt\raisebox{-0.8pt}{\scalebox{0.75}{\mbox{\rotatebox{45}{$\diamond$}}}} 
\alpha} \VEC{\eta}_{l \to \alpha}^{(k)}(\sigma_l)\mbox{ is minimal}} \\
    \!\!\!\! & {\displaystyle\mbox{decoding output:}} && \VEC{M}^{(i, 
\pm)} = {\TS\frac{1 \mp 1}{2}} \VEC{e}_i + \sum_{\alpha}^{i 
\circ\hskip-1.25pt-\hskip-1.7pt\raisebox{-0.8pt}{\scalebox{0.75}{\mbox{\rotatebox{45}{$\diamond$}}}} 
\alpha} \VEC{\mu}_{\alpha \to i}^{(n_{\rm iter} - 1/2)}(\pm)
  \end{align*}
 Here $\VEC{e}_j$ is the vector whose all but one components are equal 
to $0$, while its $j^{\rm \,th}$ component is equal to $1$. At the 
beginning of the decoding there are ``no messages'' to bits, \IE, 
$\VEC{\mu}_{\alpha \to j}^{(-1 / 2)}(\pm) = \VEC{0}$ --- the zero 
vector.

Another needed vector is $\VEC{N}^{(i)}$ that contains the 
total number bits' copies in the CT based on bit $i$:
  \begin{align*}
    \VEC{\mu}_{\alpha \to j}^{(-1 / 2)} &= \VEC{0} &
    \VEC{\eta}_{j \to \alpha}^{(k)} &= \VEC{e}_j + \sum_{\beta \ne 
\alpha}^{j 
\circ\hskip-1.25pt-\hskip-1.7pt\raisebox{-0.8pt}{\scalebox{0.75}{\mbox{\rotatebox{45}{$\diamond$}}}} 
\beta} \VEC{\mu}_{\beta \to j}^{(k - 1/2)} \NN \\
    \VEC{\mu}_{\alpha \to j}^{(k + 1 / 2)} &= \sum_{l \ne j}^{l 
\circ\hskip-1.25pt-\hskip-1.7pt\raisebox{-0.8pt}{\scalebox{0.75}{\mbox{\rotatebox{45}{$\diamond$}}}} 
\alpha} \VEC{\eta}_{l \to \alpha}^{(k)} &
    \VEC{N}^{(i)} &= 
\VEC{e}_i + \sum_{\alpha}^{i 
\circ\hskip-1.25pt-\hskip-1.7pt\raisebox{-0.8pt}{\scalebox{0.75}{\mbox{\rotatebox{45}{$\diamond$}}}} 
\alpha} \VEC{\mu}_{\alpha \to i}^{(n_{\rm iter} - 1/2)} \NN
  \end{align*}

From now on it is assumed that the transmission channel is the AWGN one, 
\IE, $h \propto (1 - \xi)$.

Consider the case when the iterative decoding produces an error on bit 
$i$, \IE, $m^{(n_{\rm iter})}_i \le 0$. Within ${\cal X}_i$ consider a 
sphere ${\cal S}_i$ with the center at $\VEC{\Xi}_0 \DEF {\rm 
CT}_i(\VEC{\xi}_0)$ and the radius $R_i \DEF \Vert \VEC{\Xi}^{(i, +)} - 
\VEC{\Xi}_0 \Vert$, where $\VEC{\Xi}^{(i, +)} \DEF \VEC{1} - 
{\VEC{\Sigma}^{(i, +)}}$, and the vector $\VEC{1}$ has all components 
being equal to $1$. All pseudo-codewords on the ${\rm CT}_i$ that have 
``$+$'' at the center are outside of or on the sphere (as 
$\VEC{\Sigma}^{(i, +)}$ is the most probable). As $\VEC{\xi}_0$ causes a 
decoding error, the noise configuration $\VEC{\Xi}^{(i, -)} \DEF \VEC{1} 
- {\VEC{\Sigma}^{(i, -)}}$, that corresponds to the pseudo-codeword 
$\VEC{\Sigma}^{(i, -)}$, is inside or on the sphere $\mathcal{S}_i$.

For brevity the mentioning of bit $i$ will be dropped from the notation 
${\rm CT}_i$, $\VEC{N}^{(i)}$, $\VEC{\Sigma}^{(i, \pm)}$, $R_i$, \ETC 
--- the bit $i$ is fixed until we find the lower weight noise 
$\VEC{\xi}$ with their knowledge. Such a procedure is repeated for all 
the bits $1 \le i \le N$ on which the decoding is in error, and the 
lowest weight noise configuration found is kept.

The set of noise configurations on ${\rm CT}$ that are closer to 
$\VEC{\Xi}^{(-)}$ than to the sphere $\mathcal{S}$ is the interior of a 
prolate spheroid with $\VEC{\Xi}_0$ and $\VEC{\Xi}^{(-)}$ as its foci.

The task is to find the admissible point $\VEC{\Xi} = {\rm 
CT}(\VEC{\xi})$ on the spheroid with the smallest weight $\Vert 
\VEC{\xi} \Vert^2$. Let $\VEC{\Xi}_\mathcal{S}$ be the point on the 
sphere ${\cal S}$ which is the closest to ${\rm CT}(\VEC{\xi})$. As 
$\VEC{\Xi}_0 = {\rm CT}(\VEC{\xi}_0)$ is the center of the sphere, we 
have ${\rm CT}(\VEC{\xi}) = (1 - \tau) {\rm CT}(\VEC{\xi}_0) + \tau 
\VEC{\Xi}_{\cal S}$ with $0 \le \tau \le 1$, so $\VEC{\Xi}_{\cal S}$ is 
actually admissible and is generated by some noise $\VEC{\xi}_{\cal S}$, 
\IE, $\VEC{\Xi}_{\cal S} = {\rm CT}(\VEC{\xi}_{\cal S})$.

Let us minimize the weight $\Vert \VEC{\xi} \Vert^2$ adding the two 
constraints $\Vert \VEC{\Xi}_\mathcal{S} - \VEC{\Xi}_0 \Vert = \Vert 
\VEC{\Xi}^{(+)} - \VEC{\Xi}_0 \Vert$ and $\Vert \VEC{\Xi}_{\cal S} - 
\VEC{\Xi} \Vert = \Vert \VEC{\Xi}^{(-)} - \VEC{\Xi} \Vert$ using 
Lagrange multipliers $\lambda$ and $\mu$, respectively:
 \begin{align*}
   \mathcal{F}(\VEC{\xi}) \DEF \Vert \VEC{\xi} \Vert^2 & + \lambda 
\bigl( \Vert \VEC{\Xi} - \VEC{\Xi}_0 \Vert^2 - \tau^2 R^2 \bigr) \\
   & + \mu \bigl( \Vert \VEC{\Xi} - \VEC{\Xi}^{(-)} \Vert^2 - (1 - 
\tau)^2 R^2 \bigr) \\
   R^2 = \sum_{k = 1}^N N_k \xi_{0k}^2 &+ 4 \sum_{k = 1}^N M^{(+)}_k (1 
- \xi_{0k})
 \end{align*}
 Here the vector $\VEC{\xi}_\mathcal{S}$ is already substituted by 
$\bigl( \VEC{\xi} - (1 - \tau) \VEC{\xi}_0 \bigr) / \tau = \VEC{\xi}_0 + 
(\VEC{\xi} - \VEC{\xi}_0) / \tau$. Varying over $\tau$ tests different 
``latitudes'' on the spheroid and gives $\lambda \tau = \mu (1 - \tau)$. 
Optimizing the quadratic function of $\VEC{\xi}$, one gets ($\gamma \DEF 
\tau / \mu$)
 \BEA
   && \xi_k = \frac{\lambda N_k \xi_{0k} + 2 \mu M^{(-)}_k} {(\lambda + 
\mu) N_k + 1} = \frac{(1 - \tau) N_k \xi_{0k} + 2 \tau M^{(-)}_k} {N_k + 
\gamma} \NN
 \EEA
 In order to satisfy the two constraints, one must have
 \begin{align*}
   & \sum_{k = 1}^N N_k (\xi_k - \xi_{0k})^2 = \tau^2 R^2 \\
   & \sum_{k = 1}^N N_k \xi_k^2 + 4 \sum_{k = 1}^N M^{(-)}_k (1 - \xi_k) 
= (1 - \tau)^2 R^2
 \end{align*}
 Subtracting one equation from another one gets a linear condition on 
$\VEC{\xi}$ (and consequently on $\tau$), which allows to express the 
parameter $\tau$ through the rest:
 \begin{align*}
   \tau = \frac{\DS \sum_{k = 1}^N \biggl[ \frac{2 \gamma N_k 
\xi_{0k}^2}{N_k + \gamma} + 4 M^{(+)}_k (1 - \xi_{0k}) - 4 M^{(-)}_k 
\Bigl( 1 - \frac{N_k \xi_{0k}}{N_k + \gamma} \Bigr) \biggr]}
 {\DS \sum_{k = 1}^N \biggl[ \frac{2 \gamma N_k \xi_{0k}^2}{N_k + 
\gamma} + 8 M^{(+)}_k (1 - \xi_{0k}) - 8 M^{(-)}_k \frac{M^{(-)}_k - N_k 
\xi_{0k}}{N_k + \gamma}\biggr]}
 \end{align*} The only remained parameter to find is $\gamma$, as $\tau$ 
is readily exppressed trough it. The task of finding the lowest weight 
noise $\VEC{\xi}$ on the spheroid is reduced to finding the root of a 
single variable function.

In practice we expect a hypothetical instanton search procedure to 
quickly end up with $m^{(n_{\rm iter})} = 0$, which means that 
$\VEC{\Xi}^{(-)}$ lies on the sphere $\mathcal{S}$. The spheroid then 
degenerates into a segment connecting $\VEC{\Xi}_0$ and 
$\VEC{\Xi}^{(-)}$. Unless $\VEC{\Xi}^{(-)} = \VEC{1} - 
\VEC{\Sigma}^{(-)}$ is admissible, the optimization will end up in 
$\VEC{\xi} = \VEC{\xi}_0$, \IE, no progress will be made. Such an 
unwanted situation is a result of dealing with the sphere $\mathcal{S}$, 
as we don't know where other pseudo-codewords on ${\rm CT}$ with ``+'' 
value at the cental bit could lie --- by dynamic programming we only did 
obtain $\VEC{\Sigma}^{(+)}$.

\end{document}